\documentclass[aps,prb,twocolumn,groupedaddress,showpacs,superscriptaddress,amssymb,amsmath]{revtex4-2}
\usepackage{graphicx}
\usepackage{dcolumn}
\usepackage{bm}
\usepackage{hyperref}
\usepackage{cleveref}
\hypersetup{
    colorlinks=true,
    linkcolor=blue,
    urlcolor=blue,
	citecolor=blue
}
\usepackage{comment}
\usepackage{color}
\usepackage[utf8]{inputenc}
\usepackage{graphicx}
\usepackage{tabularx}
\usepackage{xcolor}
\usepackage{amsmath}
\usepackage{dcolumn}
\usepackage{hyperref}
\usepackage{bm}
\usepackage{epsf}
\usepackage{braket}
\usepackage{tensor}
\usepackage{soul}

\begin{document}
\newcolumntype{M}[1]{>{\centering\arraybackslash}m{#1}}

\title{Higher-order gap ratios of singular values in open quantum systems}

\author{S. Harshini Tekur}
\email{harshini.tekur@acads.iiserpune.ac.in} 
\affiliation{Department of Physics, Indian Institute of Science Education and Research, Pune 411008, India}

\author{M. S. Santhanam}
\email{santh@iiserpune.ac.in}
\affiliation{Department of Physics, Indian Institute of Science Education and Research, Pune 411008, India}

\author{Bijay Kumar Agarwalla}
\email{bijay@iiserpune.ac.in}
\affiliation{Department of Physics, Indian Institute of Science Education and Research, Pune 411008, India}

\author{Manas Kulkarni}
\email{manas.kulkarni@icts.res.in} 
\affiliation{International Centre for Theoretical Sciences, Tata Institute of Fundamental Research,
Bangalore 560089, India}

\date{\today} 

\begin{abstract}
Understanding open quantum systems using information encoded in its complex eigenvalues has been a subject of growing interest. 
In this paper, we study higher-order gap ratios of the singular values of generic open quantum systems. We show that $k$-th order gap ratio of the singular values of an open quantum system can be connected to the nearest-neighbor spacing ratio of positions of classical particles of a 
harmonically confined log-gas 
with inverse temperature $\beta'(k)$ where $\beta'(k)$ is an analytical function that depends on $k$ and the Dyson's index $\beta=1,2,$ and $4$ that characterizes the properties of the associated Hermitized matrix. Our findings are crucial not only for understanding long-range correlations between the eigenvalues but also provide an excellent way of distinguishing different symmetry classes in an open quantum system. To highlight the universality of our findings, we demonstrate the higher-order gap ratios using different platforms such as non-Hermitian random matrices, random dissipative Liouvillians, Hamiltonians coupled to a Markovian bath, and Hamiltonians with in-built non-Hermiticity. 
\end{abstract}

\maketitle

{\textit{Introduction:}}
Understanding and characterizing various aspects of open quantum systems has been a subject of growing interest both theoretically and experimentally \cite{carmichael2009open, Haake_2001, schomerus2005quantum, OQS_Beruer, bender2007making, rotter2015review,  koch2016controlling,PhysRevA.93.062114, PhysRevA.105.032208}. Generic open quantum systems are associated with non-Hermitian matrices whose complex eigenvalues encode information about the underlying setup. Random Matrix Theory (RMT) has proved indispensable in the study of spectral correlations in such systems \cite{ Haake_2001, beenakker1997random, FEINBERG1997579, Garcia-NHCRM2002, mehta2004random, 8a873437-53d6-348d-9b7a-19c7327502a1, berkolaiko2012universality,  PhysRevLett.111.124101, schomerus2017random,Lange_Timm_2021,PRR_Ryuichi}. For example, if the correlations between eigenvalues of a system are similar to those of non-Hermitian random matrices, then the system is considered to be chaotic \cite{Haake_regular_chaotic, GrobePRL1989, prosen_PRL_2019,complex_prosen,PhysRevA.109.L050201}. On the other hand, if the eigenvalues of associated matrices have Poisson statistics, the system is considered to be localized \cite{PhysRevLett.123.090603, prosen_PRL_2019,complex_prosen,PhysRevA.109.L050201}. Such complex spectral analysis often helps in predicting several other important aspects such as nature of quantum transport and spread of wave packets \cite{PhysRevB.102.245147, PhysRevResearch.3.013208, PhysRevB.105.024303, PhysRevB.106.064205,wave_packet_non_Hermitian_Spring, hatano2024quantumtransport}. For Hermitian systems, the spectral statistics of various symmetry classes are well-studied and quite distinct \cite{brody1973statistical, berry1977level, bohigas1984characterization, GUHR1998189, PhysRevE.73.036201, oganesyan2007localization, PhysRevLett.110.084101, Atas_2013, PhysRevE.97.062212, PhysRevE.101.012216, PhysRevE.101.022222,PhysRevE.103.012208, Casal_Munoz_Molina_2021}. In the case of generic open quantum systems, there are various measures extracted from complex eigenvalues of the underlying non-Hermitian matrices \cite{ginibre1965statistical, PhysRevLett.60.1895, PhysRevLett.67.941, PhysRevLett.79.557, PhysRevLett.79.1797, PhysRevA.105.L050201, PhysRevResearch.5.033196, Mak_Bhaseen_Pal_2024,   prosen_PRL_2019,complex_prosen, PhysRevA.110.032220, PhysRevB.109.174201}. Often certain diagnostics of complex spectra yield very similar behaviour, thereby making it difficult to decode the salient features of the underlying setup. A notable example is that of the level spacing \cite{GrobePRL1989, Haake_regular_chaotic,complex_prosen} and complex spacing ratios \cite{PhysRevE.105.044144,  supp,complex_prosen} of widely different symmetry classes of non-Hermitian random matrix theory \cite{ginibre1965statistical,PhysRevB.55.1142, Bernard2002, PhysRevResearch.2.023286, NH_Jacobus, PhysRevX.13.031019,  PRXQuantum.4.030328, PhysRevResearch.5.033043}. Regardless of whether the symmetry belongs to the Ginibre Orthogonal, Unitary or Symplectic Ensembles (GinOE, GinUE, GinSE respectively) \cite{byun2023proga, byun2023progb}, the distributions for the level spacing \cite{GrobePRL1989,Haake_regular_chaotic,PhysRevLett.79.557, PhysRevLett.123.234103, prosen_PRL_2019,NHMBLHKU2019,  akemann2024transitionscomplexeigenvaluestatistics} and complex spacing ratios \cite{Jaiswal_2019, complex_prosen,PhysRevLett.124.100604,Yusipov_chaos,Scipost_Dukelsky, GGK2022, NH_Jacobus,NHMBL_Shih,PRR_Ryuichi} are characterized by the well-known cubic repulsion \cite{supp}.

Interestingly, a recent study \cite{svd_Zhenyu} has shown that the singular values (eigenvalues of $\sqrt{M^{\dagger} M}$ with $M$ being a generic non-Hermitian matrix) of underlying non-Hermitian matrices can markedly distinguish different non-Hermitian symmetry classes for both the level spacing and the gap ratios \cite{chenu_svd, mendezbermudez2024singularvaluestatisticsdirectedrandom}. However, these quantities encode information of only short-range correlations. In contrast, to understand long-range spectral correlations, a natural generalization would be the $k$-th order spacing ratio of the singular values. Although this has been studied in the context of real eigenvalues associated with Hermitian quantum systems \cite{Harshini_higher_order, Rao_Vyas_Chavda_2020, PhysRevB.102.054202, PhysRevE.107.024132}, Poisson point processes \cite{PhysRevE.73.047202}, and empirical correlation matrices \cite{bhosale2018scaling}, so far nothing is understood for singular values emerging from generic non-Hermitian systems. In addition to encoding important information about long-range correlations, this quantity is robust to details of non-universal features, such as the density of states, which is often cumbersome to factor out in other long-range quantifiers such as the singular value spectral form factor $\sigma{FF}$ \cite{chenu_svd, nandy2024}. 

In this work, we analyze the singular values $\lambda_i$ (real and non-negative) emerging from the underlying non-Hermitian matrices through the $k$-th order  non-overlapping spacing ratio $r_i^{(k)}$ defined as
\begin{equation}
    r_i^{(k)} = \frac{\lambda_{i+2k} -\lambda_{i+k}}{\lambda_{i+k}-\lambda_{i}}, \quad i,k =1,2,3, \cdots
    \label{k-th-ratio}
\end{equation}
We consider scenarios in which the singular values remain in the bulk, {\it i.e.}, $1 \ll k \ll N$ where $N$ is the number of singular values. The non-Hermitian matrices considered here have different origins: (i) non-Hermitian random matrix ensembles, (ii) random dissipative Liouvillians, (iii) Hamiltonian systems that are coupled to Lindblad bath, and finally (iv) Hamiltonians with in-built non-Hermiticity. In all cases, we find remarkable universal connection between the $k$-th order gap ratios of singular values with that of nearest-neighbor
particle spacing ratio in a one-dimensional classical gas of logarithmically interacting and harmonically confined particles \cite{8a873437-53d6-348d-9b7a-19c7327502a1, Dyson_1962_1, Dyson_1962_2} with effective inverse temperature $\beta'(k)$ given by 
\begin{equation}
\beta'(k)= \frac{k(k+1)}{2}\beta + (k-1), \quad k\geq 1.
\label{scaling}
\end{equation}
Here the Dyson index $\beta=1,2, 4$ corresponds to the symmetry classes of the singular values of the underlying non-Hermitian matrix \cite{dyson1962threefold, Atas_2013}. In other words, if the position of $N$ confined particles are $x_i$, with $i=1,2 \dots N$, then the distribution of the nearest neighbour spacing ratio of the position of particles $(x_{i+2}-x_{i+1})/(x_{i+1}-x_{i})$ drawn from the joint probability distribution 
\begin{equation}
\mathcal{P}_{\beta'(k)}(\{x_i\}) \propto \exp\Bigg[-\frac{1}{2} \sum_{i=1}^{N} x_i^2- \frac{\beta'(k)}{2} \!\! \sum_{i\neq j=1}^{N} \log \big|x_i-x_j\big|\Bigg],
\label{log-gas}
\end{equation}
matches with that of $k$-th order spacing ratio. In Eq.~\eqref{log-gas},  $\beta'(k)$ is given by Eq.~\eqref{scaling}.
Mathematically, this can be written as 
\begin{equation}
P^{(k)}(r,\beta) = P(r,\beta'(k)),
\label{scaling-distribution}
\end{equation} 
where $P(r, \beta'(k))$ is the standard nearest-level spacing ratio distribution and is given as \cite{Harshini_higher_order}
\begin{equation}
 P\big(r,\beta'(k)\big)= C_{\beta'(k)} \, \frac{(r + r^2)^{\beta'(k)}}{(1+r + r^2)^{1+ \frac{3}{2}\beta'(k)}}
 \label{P_r_beta}
\end{equation}
with  $\beta'(k)$ is given in Eq.~\eqref{scaling} and 
\begin{equation}
     C_{\beta'(k)}= \frac{3^{3(1+\beta'(k))/2} \, \Gamma(1+\beta'(k)/2)^2}{2 \pi \,\Gamma(1+\beta'(k))},
\end{equation}
where $\Gamma(x)$ is the Gamma function. While a rigorous derivation of the elegant expression in Eq.~\eqref{scaling} has been elusive so far, the numerical evidence presented in this work is highly compelling and beyond doubt.  Note that for $\beta=1,2,$ and $4$, the effective $\beta'(k)$ in  Eq.~\eqref{scaling}  is a positive integer. One can envisage this as 
a subset of the Gaussian $\beta$ ensembles \cite{10.1063/1.1507823, PhysRevLett.122.180601}. To showcase the remarkable universality of our result, we now present and discuss persuasive  numerical evidence for various cases that give rise to non-Hermitian matrices. 

\begin{figure}[t]
    \centering
    \includegraphics[width=0.95\linewidth]{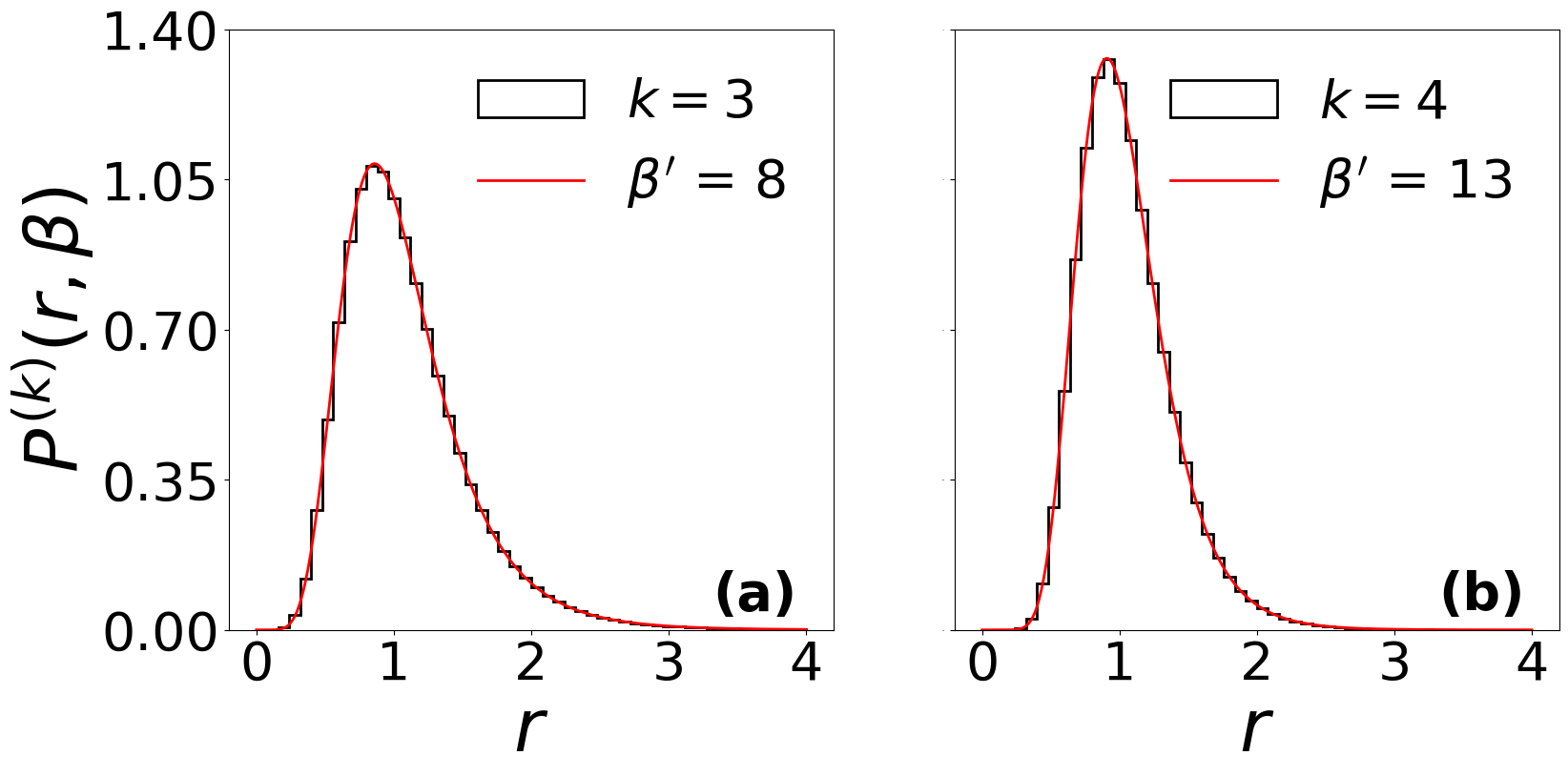}
    \caption{Plots for $k$-th order level spacing ratios [Eq.~\eqref{k-th-ratio}] for non-Hermitian random matrix (NHRM) that belongs to GinOE class. (a) and (b) represents $k=3$ and $k=4$, respectively (black-solid). We notice remarkable agreement between the higher order level spacing ratio and the nearest neighbour spacing ratio of Dyson's log gas (red-solid) with an effective inverse temperature given in Eq.~\eqref{scaling} for $\beta=1$ and holds for any $k\geq 1$. We choose NHRM of size $10^4 \times 10^4$ and obtain the statistics over 500 realizations. We further report the value of KL divergence [Eq.~\eqref{KL-divergence}] to be $0.0088$ and $0.0005$, for (a) and (b), respectively, thereby cementing the remarkable agreement.}
    \label{fig:GinOE-NHRM}
\end{figure}

{\textit{Non-Hermitian random matrix ensembles:}}
\label{non-H-RMT}
We first discuss the $k$-th order level spacing ratio given in Eq.~\eqref{k-th-ratio} for standard non-Hermitian random matrix (NHRM) ensembles, namely GinOE, GinUE, and GinSE \cite{byun2023proga, byun2023progb}. In Fig.~\ref{fig:GinOE-NHRM}, the $k$-th order level spacing ratio for GinOE ensemble is displayed. The singular values of large GinOE matrices are computed. The $k$-th order level spacing ratios extracted from these singular values match perfectly with the nearest-neighbor spacing ratio of a classical log-gas with inverse temperature $\beta'(k)$ given in Eq.~\eqref{scaling} with Dyson index $\beta=1$. Furthermore, to make a thorough quantitative check, we compute the Kullback-Leibler (KL) divergence \cite{Joyce2011} which is a measure of the distance between two probability distributions $A(x)$ and $B(x)$ and is defined as
\begin{equation}
D_{KL}(A || B) = \sum_{x} A(x) \log \Big[ \frac{A(x)}{B(x)}\Big].
\label{KL-divergence}
\end{equation}
If the two distributions are identical, then $D_{KL}(A || B)=0$. In practice, vanishingly small $D_{KL}(A || B)$ indicates that $A$ and $B$ are nearly the same. In Fig.~\ref{fig:GinOE-NHRM}, KL divergence for $k=3$ and $k=4$ is reported. The negligible values demonstrate remarkable agreement between $P^{(k)}(r,\beta)$ and $P(r,\beta'(k))$ with $\beta'(k)$ given in Eq.~\eqref{scaling}.  The universality of this result for other standard random matrix ensembles such as GinUE and GinSE is provided in the Supplementary Material \cite{supp}. Moreover, in the Supplementary Material \cite{supp}, several other important classes of random matrices are discussed whose symmetries are closely connected with several physical systems, some of which are discussed later.  

\begin{figure}[t]
    \centering
    \includegraphics[width=0.95\linewidth]{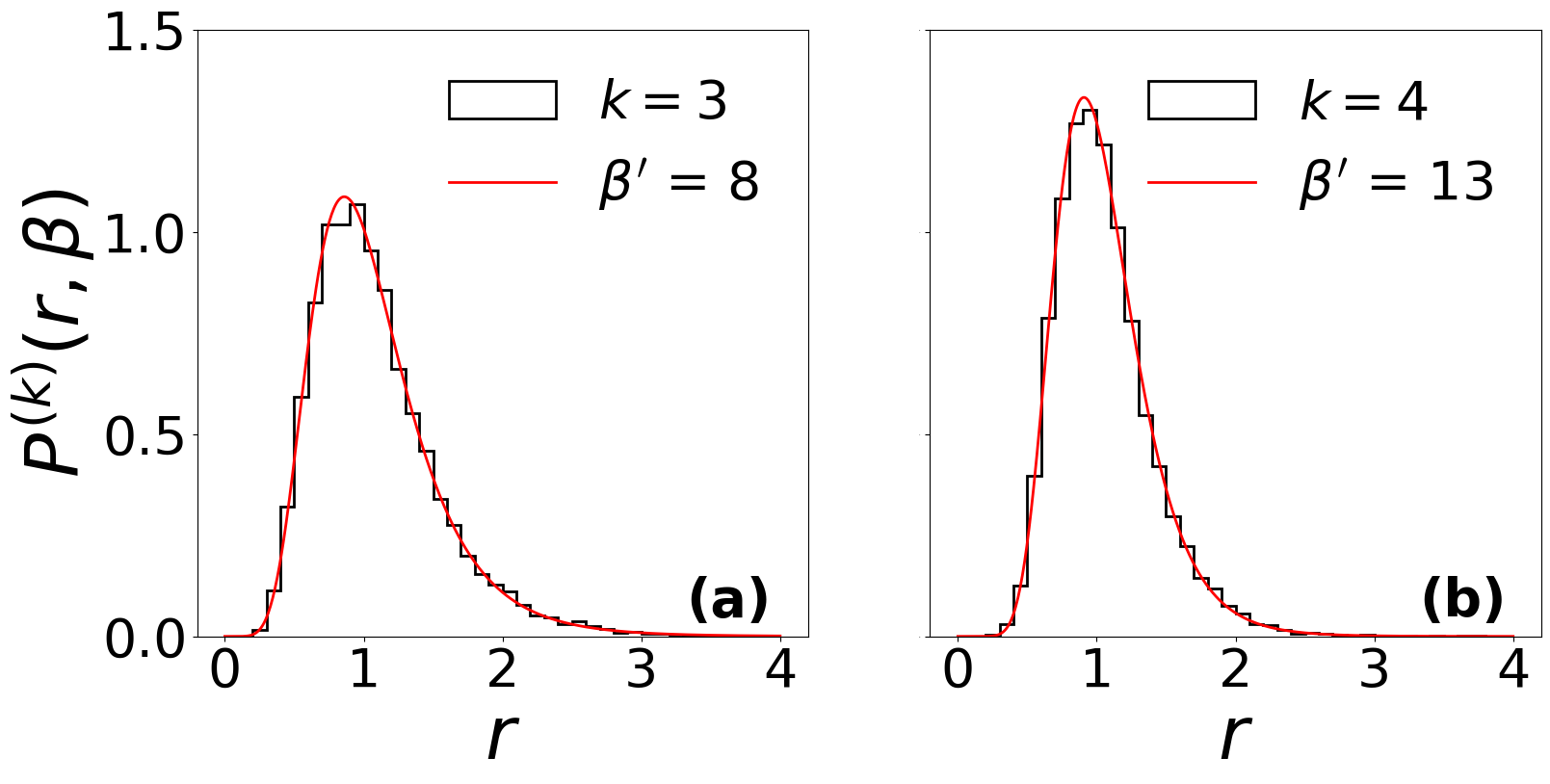}
    \caption{Plots for $k$-th order level spacing ratios [Eq.~\eqref{k-th-ratio}] for the random Liouvillian given in Eq.~\eqref{eq:RL}. Here (a) and (b) represents $k=3$ and $k=4$, respectively (black-solid). The singular value statistics follows the scaling relation [Eq.~\eqref{scaling-distribution}] for $\beta =1$ (red-solid). Similar to Fig.~\ref{fig:GinOE-NHRM}, we once again observe a remarkable agreement. The results here are obtained for system size $M=4$ for $50$ realizations. In this case we find the KL divergence [Eq.~\eqref{KL-divergence}] to be $0.020$ and $0.0019$, for (a) and (b), respectively.}
    \label{fig:rand_L}
\end{figure}

{\textit{Random Liouvillian models:}}
\label{RLM}
We now discuss a construction which describes generic open quantum systems and respects the required mathematical properties such as trace preservation and complete positivity \cite{OQS_Beruer}. Such a scenario is well-described by a Lindblad quantum master equation $\partial_t \rho_t = {\cal L}(\rho_t)$ with dynamics governed by a Liouvillian ${\cal L}$ which typically takes the Gorini-Kossakowski-Sudarshan-Lindblad form as given by $\mathcal{L}(\rho)= \mathcal{L}_H(\rho)+\mathcal{L}_D(\rho)$ where 
$\mathcal{L}_H(\rho) = -i [H, \rho]$ is the unitary part with system Hamiltonian $H$. We assume that the Hilbert space dimension is $2^M$ which is the case for say a system of $M$ qubits. $\mathcal{L}_D(\rho)$ corresponds to the purely dissipative part given by
\begin{equation}
    \mathcal{L}_D(\rho) = \sum_{i,j=1}^{2^{2M}-1} K_{ij} \Big[L_i\rho L_j^\dag - \frac{1}{2}\{L_j^\dag L_i,\rho\}\Big].
    \label{eq:RL}
\end{equation}
Here $L_i, i=1,2, \cdots 2^{2M}\!-\!1$ is the traceless Lindblad operator that satisfy the orthonormality condition ${\rm Tr}\big[L_i L_j^{\dagger}\big]= \delta_{ij}$ \cite{OQS_Beruer}. The matrix $K$ is chosen to be random \cite{PhysRevLett.123.140403,Can_2019, PhysRevLett.124.100604,Sa_2020,10.21468} and always satisfies the positive semi-definite property. This guarantees that the time evolution of the density matrix always respects the positivity condition. Furthermore, in our setup we assume that there is no Hamiltonian, i.e., we set $H=0$ and analyze the $k$-th order level spacing ratio only in presence of the dissipative part [Eq.~\eqref{eq:RL}]. We consider a setup of $M$ qubits which allows $2^{2M}-1$ possible $L_i$ operators consisting of Pauli strings. In other words, each $L_i$ is a direct product of Pauli matrices $\sigma$'s or identity matrix $I_{2\times 2}$ excluding the situation where every element is an identity. More concretely, each $L_i$ is of the form $\frac{1}{2^{M/2}} \sigma_{x_1} \otimes \sigma_{x_2}  \cdots \otimes \sigma_{x_n}, x_i \in 0,1,2,3$ where $0,1,2,3$ represents $I_{2\times 2}$, and the Pauli matrices $\sigma_x$, $\sigma_y$, and $\sigma_z$, respectively. The random matrix $K$ is obtained as follows: we first prepare a $(2^{2M}\!-\!1) \times (2^{2M}\!-\!1)$ diagonal matrix $D$ whose entries are chosen from by sampling from a uniform box distribution $[0,1]$. We then rotate this diagonal matrix with a random unitary matrix $U$ sampled from the Haar measure \cite{mezzadri2007} to give the required $K=U^{\dagger} D U$ matrix. Given the $K$ matrix and all possible $L_i$ operators, we construct the Liouvillian in a matrix form of dimension $2^{2M} \times 2^{2M}$ and compute its singular values. In Fig.~\ref{fig:rand_L}, we plot the the $k$-th order level spacing ratios for $k=3$ and $k=4$. The Liouvillian constructed in this example falls under $AI^{\dagger}$ symmetry class. Interestingly, we find that the corresponding $k$-th order spacing ratio for the singular values match remarkably well with the nearest neighbour spacing ratio of Dyson's log-gas with an effective inverse temperature given by Eq.~\eqref{scaling} with Dyson index $\beta=1$. It is interesting and important to see whether such observations hold in physical setups where Hamiltonians are coupled to Markovian baths. This is what we address next. 

\begin{figure}[t]
    \centering
    \includegraphics[width=0.95\linewidth]{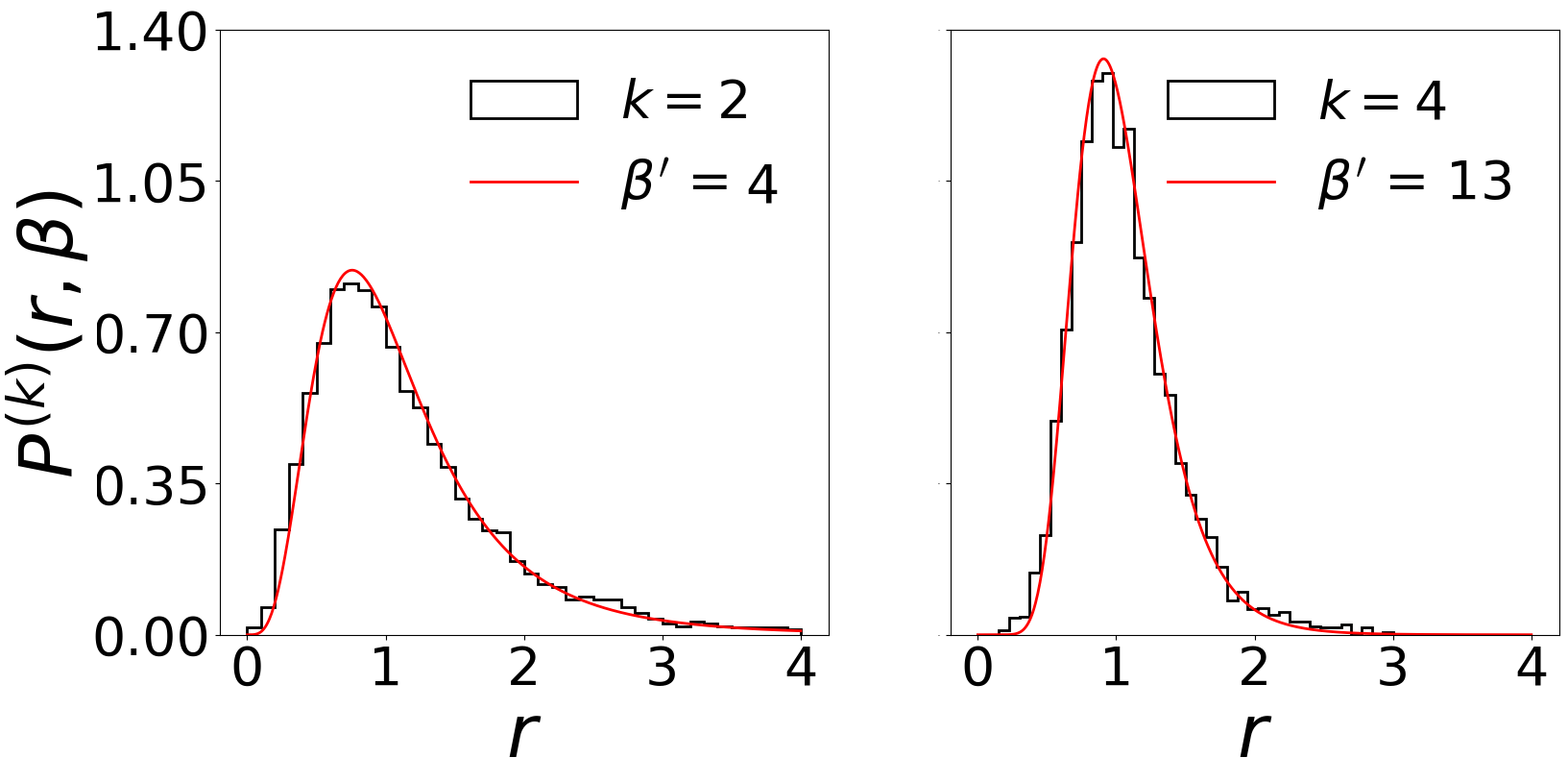}
    \caption{Plots for $k$-th order level spacing ratios [Eq.~\eqref{k-th-ratio}] for the physical Lindbladian given in Eq.~\eqref{eq:PL} with Hamiltonian given in Eq.~\eqref{spin-ising} and the jump operators given by $L_i = \sqrt{\gamma} \, \sigma_i^{-}$. Here (a) and (b) represents $k=2$ and $k=4$, respectively (black-solid). Similar to Fig.~\ref{fig:GinOE-NHRM}, we once again observe a remarkable agreement with  $\beta=1$ (red-solid). The results here are obtained for system size $M=6$ averaged over $50$ different realizations. We choose $J=1$, $h_x= -1.05$, $h_z = 0.2$, and $\gamma=0.77$. In this case we find the KL divergence [Eq.~\eqref{KL-divergence}] to be $0.008$ and $0.032$, for (a) and (b), respectively.}
    \label{fig:PL}
\end{figure}

{\textit{Physical Lindbladian:}} 
We now discuss a $M$-qubit physical setup where a Hamiltonian is coupled to a Markovian bath. This open quantum system is described by the Lindblad equation $\partial_t \rho_t  = \mathcal{L} (\rho_t)$ where the Liouvillian $\mathcal{L}$ takes the standard form with unitary dynamics governed by a Hamiltonian $H$, and dissipative dynamics is governed by Lindbladians $L_i$'s that describes coupling to a Markovian bath. The Lindblad equation is given by
\begin{equation}
    \mathcal{L}(\rho)= -i[H,\rho] + \sum_{i=1}^{M} \left[L_i \rho L_i^{\dagger} - \frac{1}{2} \{L_i^{\dagger} L_i , \rho\} \right],
    \label{eq:PL}
\end{equation}
where the Hamiltonian $H$ is taken as a variant of the quantum Ising model with open boundary conditions \cite{PhysRevX.13.031019}, and is given by
\begin{equation}
    H = -J \sum_{i=1}^{M-1} \big(1+\epsilon_i\big) \, \sigma_i^z \sigma_{i+1}^z - \sum_{i=1}^M \big(h_x \, \sigma_i^x+ h_z \, \sigma_i^z\big).
    \label{spin-ising}
\end{equation}
Here $J>0$ is the nearest-neighbour coupling constant between the Pauli spins with $\epsilon_i$ being an additional onsite random disorder parameter which is sampled from a uniform distribution $[-0.1, 0.1]$. $h_x$ and $h_z$ are the constant fields in the $x$ and $z$ directions, respectively. 
The jump operators $L_i$, attached at each site, are of the form $L_i = \sqrt{\gamma}\,\sigma_i^-$, ($\sigma_i^- = \sigma_i^x -i\,\sigma_i^y$) representing the damping mechanism of spin excitation at each site with a rate $\gamma$. 
In Fig.~\ref{fig:PL}, we plot the higher-order spacing ratios of the singular values for $k=2$ and $k=4$ and observe a good agreement with the nearest-neighbor spacing ratio with the effective $\beta'(k)$ evaluated at $\beta=1$. So far in all previous cases we discussed the situation when the setup is essentially in an ergodic phase. It will therefore be interesting to consider a  situation where the underlying setup can host different phases, in particular, non-ergodic and ergodic. We therefore now analyze the $k$-th order spacing ratios for singular values of Hamiltonians with in-built non-hermiticity that contain rich phases.  

{\textit{Non-Hermitian Hamiltonian:}} We consider a $M$-site hard-core boson model with non-reciprocal hopping and a complex on-site potential. The Hamiltonian is given by
\begin{eqnarray}
    H  =  \sum_{j=1}^M &\Big[& - J \left( e^{g}\, c_j^\dag  c_{j+1} + e^{-g}\, c_{j+1}^\dag  c_j\right)  \nonumber \\
    &+& \Big(h_j + i \gamma (-1)^j \Big)  n_j + V  n_j \, n_{j+1} \Big]
    \label{eq:GinUE_H}
\end{eqnarray}
and we use periodic boundary condition. This many-body system therefore has two different origins of non-hermiticity, namely of hopping ($g$) and of on-site ($\gamma$) type. Here, $n_j=c_j^{\dagger} c_j$ is the number operator. The onsite term in Eq.~\eqref{eq:GinUE_H} consists of two parts - a random, real parameter $h_j \in [-h,h]$, and a purely imaginary parameter $\pm i\gamma$ representing alternate gain/loss. In our case, we take $g=0.1$, $V=2$, and $\gamma=0.1$. Different phases of this non-Hermitian model can be realized by tuning the disorder strength $h$. For the parameters given above, $g, V$, and $\gamma$, we find that when $h=2$, the system is deep in the ergodic regime, whereas when $h=14$, it is deep in the non-ergodic regime. Therefore, the setup in Eq.~\eqref{eq:GinUE_H} is ideally suited for studying $k$-th order spacing ratio for singular values in difference phases. 

\begin{figure}[t]
    \centering
    \includegraphics[width=0.95\linewidth]{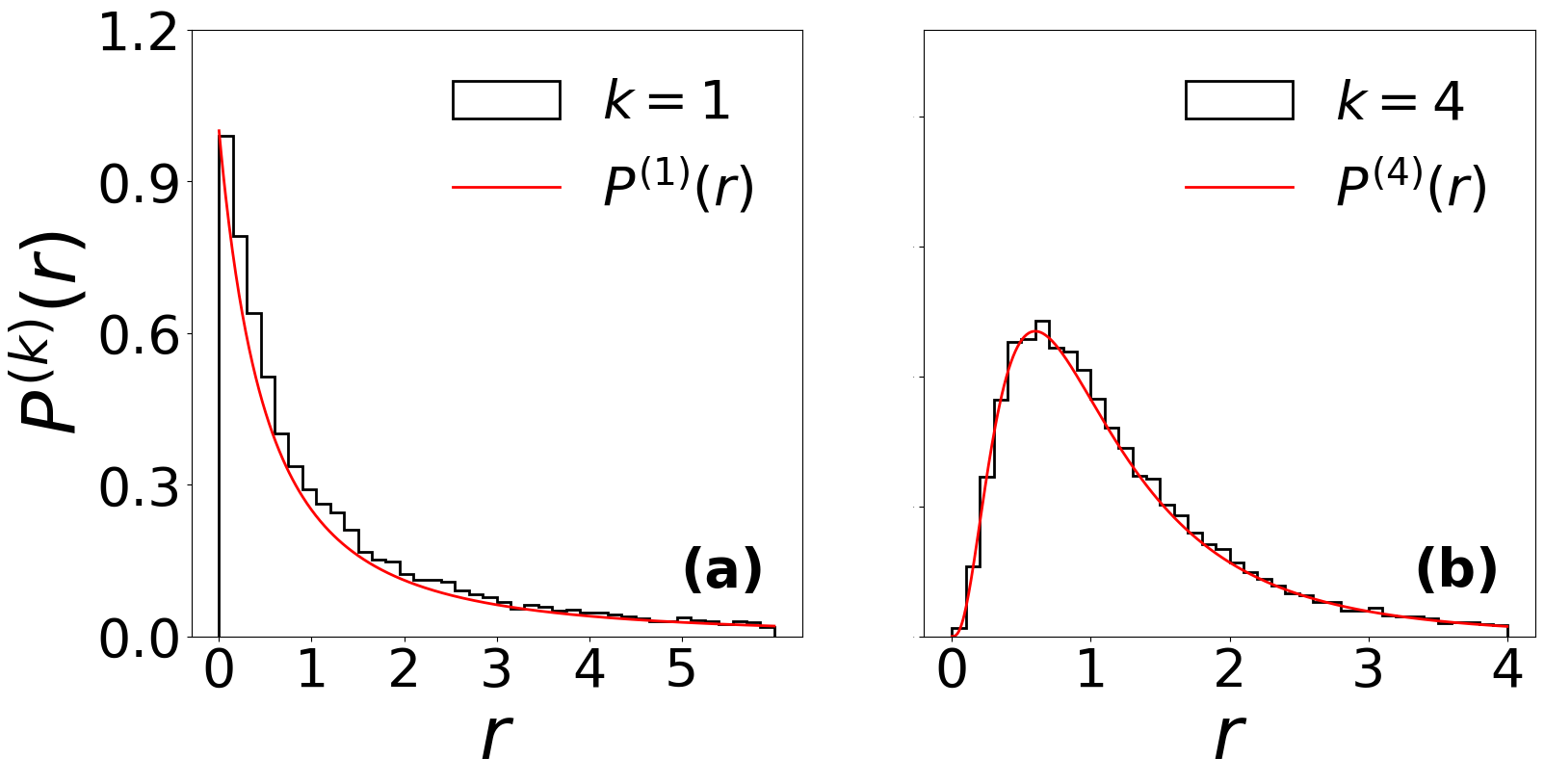}
    \caption{Plots for $k$-th order level spacing ratios [Eq.~\eqref{k-th-ratio}] for the non-Hermitian Hamiltonian [Eq.~\eqref{eq:GinUE_H}] deep in the localized regime ($h=14$). Here (a) and (b) represents $k=1$ and $k=4$, respectively (black-solid). We observe remarkable agreement with Eq.~\eqref{singular-poisson} (red-solid). The corresponding KL diverges are $0.1016$ and $0.0098$, respectively. We consider $M=16$, $J=1$, $g=0.1$, $V=2$, and $\gamma=0.1$ and the lattice is half filled. The spacing ratio distributions are obtained by performing an average over $50$ different realizations.}
    \label{fig:nhh2p}
\end{figure}

We start from the regime of high disorder ($h=14$) and study the nature of singular values. We find that the complex spectral statistics of the Hamiltonian follow 2D poisson statistics. On the other hand, the $k$-th order spacing ratios of the singular values follow the $k$-th order spacing ratios that emerges from 1D Poisson statistics and is given as \cite{symmetry_deduction}  
\begin{equation}
P^{(k)}(r)=\frac{(2k-1)!}{[(k-1)!]^2}\frac{r^{k-1}}{(1+r)^{2k}}.
\label{singular-poisson}
\end{equation}
We show remarkable agreement between the $k$-th level spacing ratios of the singular values with Eq.~\eqref{singular-poisson} in Fig.~\ref{fig:nhh2p}.

\begin{figure}[t]
    \centering
    \includegraphics[width=0.95\linewidth]{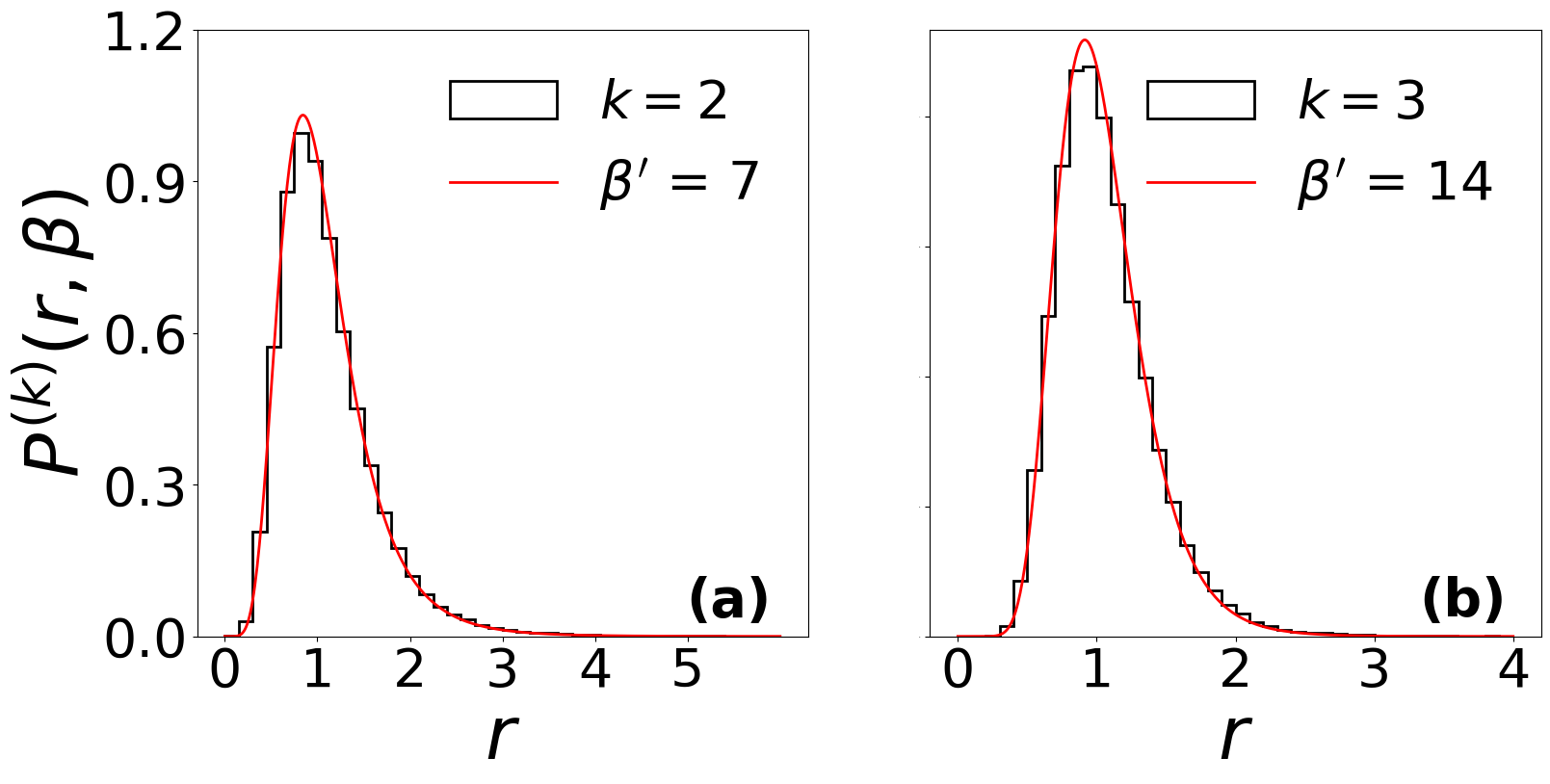}
    \caption{Plots for $k$-th order level spacing ratios [Eq.~\eqref{k-th-ratio}] for the non-Hermitian Hamiltonian [Eq.~\eqref{eq:GinUE_H}] deep in the ergodic phase ($h=2$). Here (a) and (b) represents $k=2$ and $k=3$, respectively (black-solid). The corresponding KL diverges are $0.0048$ and $0.0098$, respectively. All other parameters are same as in Fig.~\ref{fig:nhh2p}. The spacing ratio distributions are obtained by performing an average over $50$ different realizations. The numerical results are in excellent agreement with Eq.~\eqref{P_r_beta} setting $\beta=2$ (red-solid).}
    \label{fig:GinUE_H}
\end{figure}

Now we discuss the regime of low disorder strength ($h=2$) where the system is deep in the ergodic phase. The singular values of the $k$-th order level spacing ratios is once again well described by non-Hermitian random matrix theory, in particular, by the Ginibre unitary ensemble (GinUE, $\beta=2$). Recall that, when one considers complex spectra analysis, such as complex spacing distributions, it is difficult to differentiate various classes of non-Hermitian random matrix theory. For example, many of the different random matrix ensembles display the same cubic-level repulsion for complex spacing ratios \cite{supp}. The power of the singular values lie in the fact that it is uniquely suited to distinguish several non-Hermitian random matrix ensembles, as we have demonstrated here. Moreover, the $k$-th order level spacing ratio seems to be in an excellent agreement with the prediction in Eq.~\eqref{P_r_beta} with $\beta=2$. This is demonstrated in Fig.~\ref{fig:GinUE_H}.

{\textit {Summary:}} We have highlighted the immense importance of singular values of non-Hermitian random matrices through the lens of higher-order level spacing ratios. By focusing on widely different types of non-Hermitan systems, in all cases, we observe a remarkable agreement between the $k$-th order level spacing ratio of singular values and the nearest-neighbour spacing ratio of classical log-gas particles [Eq.~\eqref{log-gas}] confined in a harmonic trap at an effective inverse temperature $\beta'(k)$ through Eq.~\eqref{scaling}. This intriguing agreement therefore highlights that the use of singular value statistics is a promising route to distinguish different symmetry classes which are otherwise often difficult to demarcate using the conventional procedure of complex spectra analysis. 

In future, it will be interesting to explore a systematic correspondence between different symmetry classes \cite{svd_Zhenyu, PhysRevX.13.031019} of non-Hermitian random matrices and the corresponding singular value statistics through higher-order gap ratios and other diagnostics such as the spectral form factor \cite{PhysRevResearch.3.L012019, Spectral_random, dong2024measuringspectralformfactor,das2024proposalmanybodyquantumchaos} constructed from singular eigenvalues $\sigma{FF}$ \cite{chenu_svd,nandy2024}. The deep connection developed in this work can also be adapted to study transitions from delocalized to localized phases in non-Hermitian power-law banded random matrices \cite{plrbm_1,plrbm_2}.

{\textit {Acknowledgements:}} BKA acknowledges CRG Grant No. CRG/2023/003377 from Science and Engineering Research Board (SERB), Government of India. BKA, MSS, and SHT would like to acknowledge funding from the National Mission on Interdisciplinary  Cyber-Physical  Systems (NM-ICPS)  of the Department of Science and Technology (DST), Govt.~of  India through the I-HUB  Quantum  Technology  Foundation, Pune, India. SHT acknowledges the National Supercomputing Mission (NSM) for providing computing resources of ‘PARAM Brahma’ at IISER Pune, which is implemented by C-DAC and supported by the Ministry of Electronics and Information Technology (MeitY) and DST, Government of India.  M. K. thanks the VAJRA faculty scheme (No.~VJR/2019/000079) from SERB, Government of India. MK acknowledges support of the Department of Atomic Energy, Government of India, under Project No. RTI4001. MK thanks the hospitality of Department of Physics, IISER, Pune. BKA thanks the hospitality of International Centre of Theoretical Sciences (ICTS), Bangalore, India under the associateship program.

\bibliography{references}
\newpage

\onecolumngrid

\setcounter{equation}{0}
\setcounter{figure}{0}
\renewcommand{\theequation}{S\arabic{equation}}
\renewcommand{\thefigure}{S\arabic{figure}}
 
\begin{center}
{\textbf{\underline{Supplementary Material}}}
\end{center}

\section{Recap of quantities of interest and main findings}
\label{sec:recap}
In this section, we will recall some quantities of interest and central results that were presented in the main text. 
We investigated higher-order level spacing ratios of the singular values of generic non-Hermitian systems~\cite{svd_Zhenyu} and their random matrix counterparts.
Specifically, we study the $k$-th order, non-overlapping spacing ratio $r_i^{(k)}$, defined as
\begin{equation}
    r_i^{(k)} = \frac{\lambda_{i+2k} -\lambda_{i+k}}{\lambda_{i+k}-\lambda_{i}}, \quad i,k =1,2,3, \cdots
    \label{k-th-ratio_supp}
\end{equation}
As mentioned in the main text, we restrict ourselves to the bulk of the spectrum i.e., $1 \ll k \ll N$ where $N$ is the number of singular values. Through strong numerical evidence, we show that there is a scaling relation between $k$-th order spacing ratios given as a function of the Dyson index $\beta$ and $k$. In other words, given a value for $k$, and starting from a non-Hermitian random matrix (or system) with singular value statistics corresponding to symmetry classes with $\beta=1, 2$ or $4$, the  $\beta'(k)$ is given as 
\begin{equation}
\beta'(k)= \frac{k(k+1)}{2}\beta + (k-1), \quad k\geq 1.
\label{scaling_supp}
\end{equation}
Remarkably, our work shows that the scaling in Eq.~\eqref{scaling_supp} which was demonstrated for hermitian systems \cite{Harshini_higher_order} also holds for singular values of non-Hermitian systems.
The relation in Eq.~\eqref{scaling_supp} accurately describes the distribution of the $k$-th order ratios as 
\begin{equation}
\label{eq:P_sup}
P^{(k)}(r,\beta) = P(r,\beta'(k))\, ,
\end{equation}
where $P(r, \beta'(k))$ is the standard nearest-level spacing ratio distribution, and this is of the form
\begin{equation}
 P\big(r,\beta'(k)\big)= C_{\beta'(k)} \, \frac{(r + r^2)^{\beta'(k)}}{(1+r + r^2)^{1+ \frac{3}{2}\beta'(k)}}\,,
 \label{P_r_beta_supp}
\end{equation}
with  
\begin{equation}
     C_{\beta'(k)}= \frac{3^{3(1+\beta'(k))/2} \Gamma(1+\beta'(k)/2)^2}{2 \pi \Gamma(1+\beta'(k))},
\end{equation}
where $\Gamma(x)$ is the Gamma function. To make a comparison between the $k$-th order spacing ratio distribution and the scaled distribution [Eq.~\eqref{P_r_beta_supp}], we also calculate the Kullback-Leibler (KL) divergence, defined as
\begin{equation}
D_{KL}(A || B) = \sum_{x} A(x) \log \Big[ \frac{A(x)}{B(x)}\Big].
\label{KL-divergence_supp}
\end{equation}
In the main text, we elucidate this deep connection in various platforms  (i) non-Hermitian random matrix ensembles, (ii) random dissipative Liouvillians, (iii) Hamiltonian systems that are coupled to Lindblad bath, and finally (iv) Hamiltonians with in-built non-Hermiticity. This supplementary is devoted to present more examples to strengthen the remarkable universal aspect. In particular, we provide results for more $k$ values and for more types of non-Hermitian systems that are absent in the main text.

\begin{figure}
    \centering
    \includegraphics[width=0.4\linewidth]{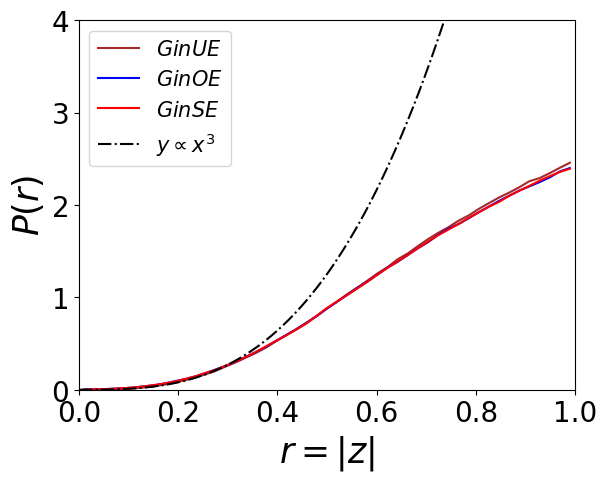}
   \caption{Plot showing the complex spacing ratio distribution given in Eq.~\eqref{eq:csr_supp} for the Ginibre Unitary (brown), Orthogonal (blue) and Symplectic (red) Ensembles for $10^4 \times 10^4$ random matrices averaged over $3500$ realizations. Also plotted is the curve for $y \propto x^3$ (black dashed-dotted line) to illustrate the cubic repulsion exhibited by all three ensembles for small values of spacing ratio $r$. It can also be seen that the  different random matrix theory curves are essentially indistinguishable.}
    \label{fig:CSR_supp}
\end{figure}

Before proceeding further, we make a few important remarks. The $k$-th order level spacing ratio of singular values contains information about long-range correlations between the eigenvalues of the system. Furthermore, as is evidenced both in main and in supplementary, it proves to be an indispensable tool for differentiating various symmetry classes of non-Hermitian matrices. This is however not the case with the standard measure of dissipative quantum systems, namely, complex spacing ratio \cite{complex_prosen}, given as 
\begin{equation}
z_k = \frac{E^{NN}_k-E_k}{E^{NNN}_k-E_k}=r_k\,e^{i\theta_k},
\label{eq:csr_supp}
\end{equation}
where $E_k$'s are the complex eigenvalues of non-Hermitian matrices. The superscript $NN$ and $NNN$ denote the nearest neighbour and next-nearest neighbour on the complex plane, respectively. In Fig. \ref{fig:CSR_supp}, we plot the distribution of $r_k$ given in Eq.~\eqref{eq:csr_supp} for Ginibre Unitary (GinUE), Orthogonal (GinOE) and Symplectic Ensembles (GinSE). We note that all three ensembles essentially have the same behaviour and display cubic repulsion for small $r$. Therefore, diagnostics such as Eq.~\eqref{eq:csr_supp} is often ill-suited for differentiating different symmetry classes. In such cases, singular values of the associated non-Hermitian random matrices prove to be instrumental. 

\begin{figure}
    \centering
    \includegraphics[width=0.95\linewidth]{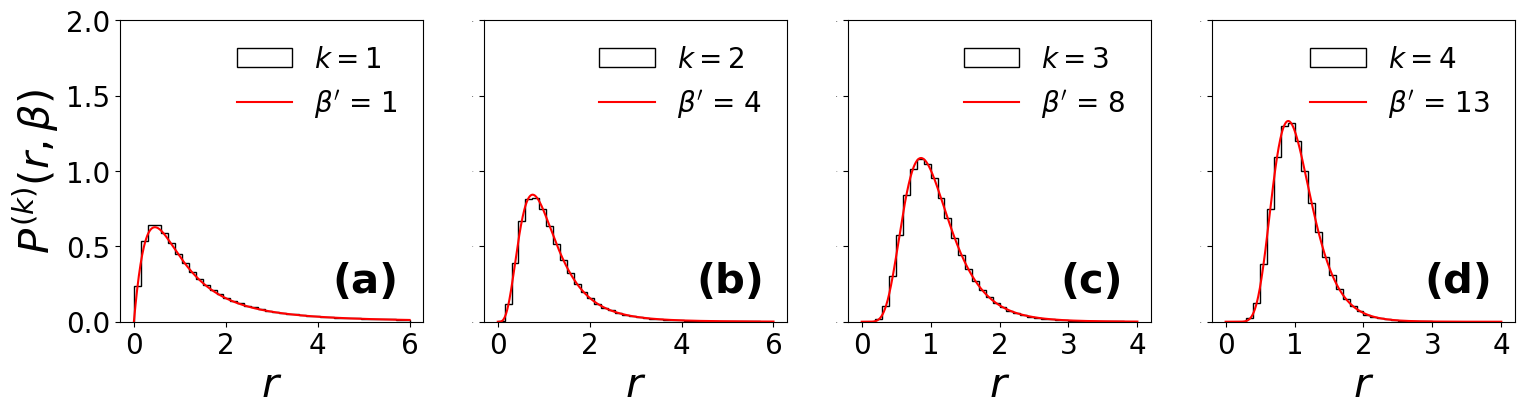}
    \caption{Plots for $k$-th order level spacing ratios [Eq.~\eqref{k-th-ratio_supp}] for non-Hermitian random matrix (NHRM) that belongs to GinOE class. (a),(b), (c) and (d) represent $k=1$, $k=2$, $k=3$ and $k=4$, respectively (black-solid). We notice remarkable agreement between the higher-order level spacing ratio and the nearest neighbour spacing ratio of Dyson's log gas with an effective inverse temperature given $\beta'(k)$ in Eq.~\eqref{scaling_supp} for $\beta=1$ and holds for any $k\geq 1$ (red-solid). We choose NHRM of size $10^4 \times 10^4$ and obtain the statistics over 100 realizations. We further report the value of KL divergence to be $0.0008$, $0.0031$, $0.0018$ and $0.0023$, for (a), (b), (c) and (d) respectively, thereby cementing the remarkable agreement.}
    \label{fig:GinOE-NHRM_supp}
\end{figure}

\section{$k$-th level spacing ratio in Non-Hermitian Random Matrix Ensembles}
\begin{figure}
    \centering
    \includegraphics[width=0.95\linewidth]{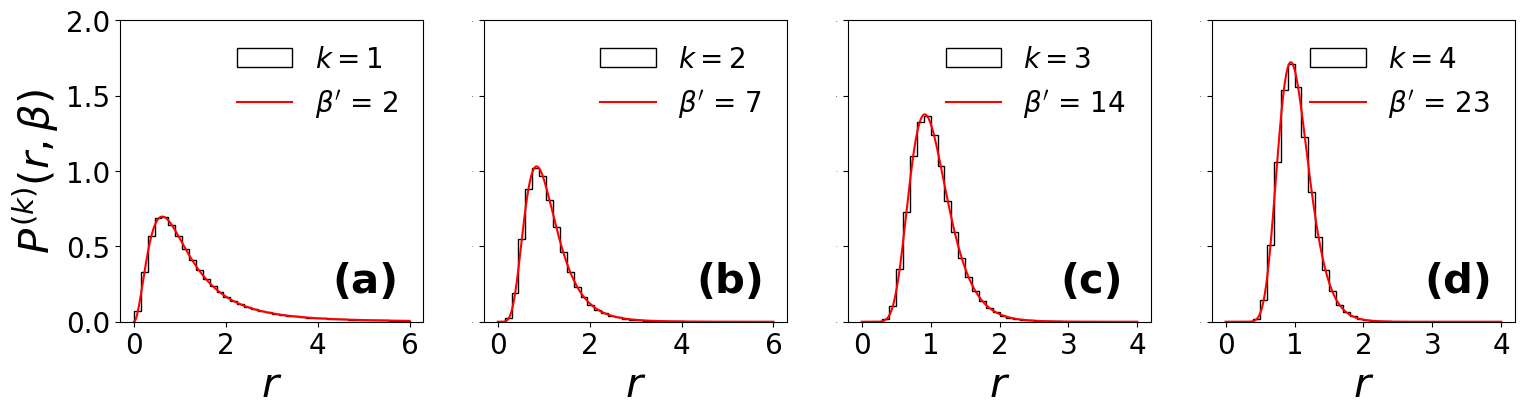}
   \caption{Plots for $k$-th order level spacing ratios [Eq.~\eqref{k-th-ratio_supp}] similar to the one shown in Fig.~\ref{fig:GinOE-NHRM} but for GinUE class ($\beta =2$) (black-solid). We choose NHRM of size $10^4 \times 10^4$ and obtain the statistics over 500 realizations. We further report the value of KL divergence to be $0.0042$, $0.0024$, $0.0018$ and $0.0092$, for (a), (b), (c), and (d) respectively, thereby showing excellent agreement (red-solid).}
    \label{fig:GinUE-NHRM_supp}
\end{figure}

\begin{figure}
    \centering
    \includegraphics[width=0.95\linewidth]{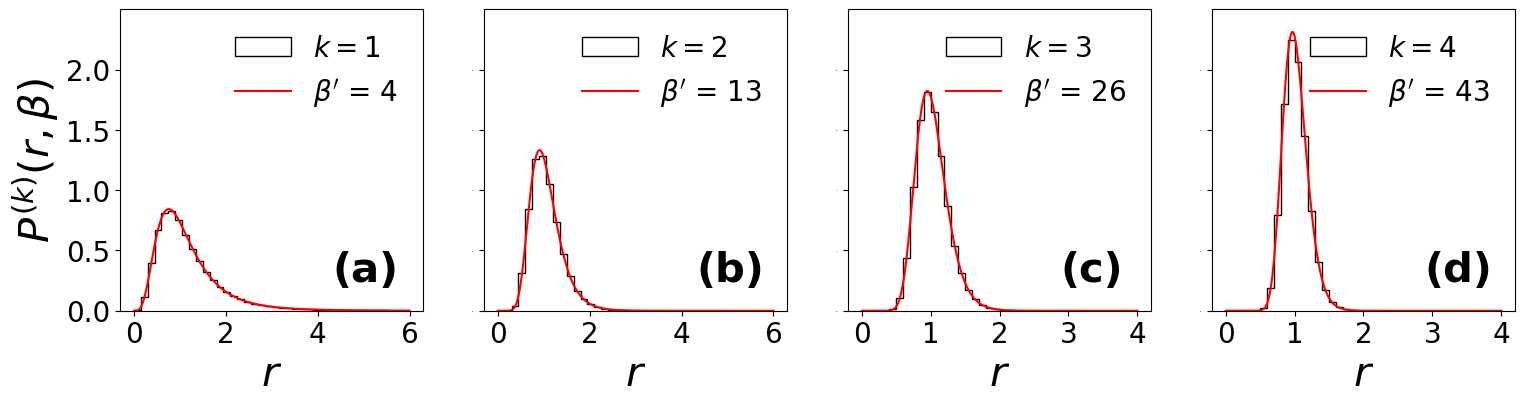}
    \caption{Plots for $k$-th order level spacing ratios [Eq.~\eqref{k-th-ratio}]
    similar to the one shown in Fig.~\ref{fig:GinOE-NHRM_supp} but for GinSE class (black-solid). We choose NHRM of size $10^4\times 10^4$ and obtain the statistics over 100 realizations. We notice remarkable agreement between the higher-order level spacing ratio and the nearest neighbour spacing ratio of Dyson's log gas with an effective inverse temperature given $\beta'(k)$ in Eq.~\eqref{scaling_supp} for $\beta=4$ (red-solid). We further report the value of KL divergence to be $0.0031$, $0.0022$ $0.0051$ and $0.0093$, for (a), (b), (c) and (d), respectively, thereby confirming excellent agreement.}
    \label{fig:GinSE-NHRM_supp}
\end{figure}

We calculate higher order spacing ratio distributions for $k=1$ to $k=4$ for the GinOE, GinUE, and GinSE ensembles. The results are shown in Figs.~\ref{fig:GinOE-NHRM_supp}, \ref{fig:GinUE-NHRM_supp} and \ref{fig:GinSE-NHRM_supp} respectively, and we observe an excellent agreement with the scaling relation Eq.~\eqref{scaling_supp} . This is also evidenced by the values of the KL divergence [Eq.~\eqref{KL-divergence_supp}] shown in Table \ref{tab:my-table} for all three ensembles.

\begin{table}[ht]
\begin{tabular}{|c|c|l|l|l|}
\hline
$k$     & 1 & \multicolumn{1}{c|}{2} & \multicolumn{1}{c|}{3} & \multicolumn{1}{c|}{4} \\ \hline
GinOE & 0.0065  &  0.0031          &     0.0018                   & 0.0023                       \\ \hline
GinUE & 0.0042   & 0.0024          &   0.0018                     &      0.0092                  \\ \hline
GinSE & 0.0031  & 0.0022           &         0.0051               &   0.0093                     \\ \hline

\end{tabular}
\caption{Table containing the values of KL Divergence [Eq.~\eqref{KL-divergence_supp}] for GinOE, GinUE and GinSE spacing ratios for $k=1,2,3,4$.}
\label{tab:my-table}
\end{table}

\section{$k$-th level spacing ratio in Poisson distribution}
\begin{figure}[ht]
    \centering
    \includegraphics[width=0.95\linewidth]{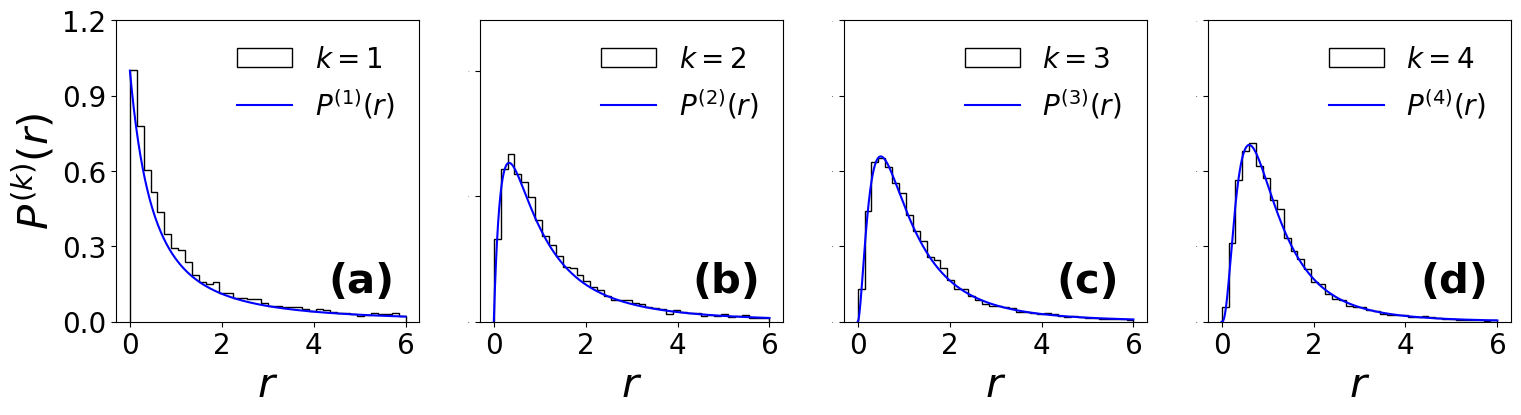}
    \caption{$k$-th order level spacing ratios from the singular-values of a 2D-Poisson process for $k=1$ to $k=4$ (black-solid). The statistics correspond to those of a 1D Poisson process (blue-solid) with the KL divergences found to be 0.1063, 0.011, 0.0043, and 0.0061 respectively.}
    \label{fig:Poisson_supp}
\end{figure}

A random matrix corresponding to a 2D Poisson process is computed by constructing a random diagonal matrix consisting of entries of the form $x+iy$ corresponding to random points $(x,y)$ scattered on a 2D plane. We then calculate its singular values and observe that the higher-order level spacing ratios for these singular values follow the scaling relation for the 1D Poisson distribution given as \cite{symmetry_deduction}:
\begin{equation}
    P^{(k)}(r)=\frac{(2k-1)!}{[(k-1)!]^2}\frac{r^{k-1}}{(1+r)^{2k}}.
\end{equation}
This is shown for $k=1$ to $k=4$ for the singular values of a $5000\times 5000$ diagonal random matrix generated from a 2D Poisson point process averaged over 50 realizations in Fig.~\ref{fig:Poisson_supp}.

\begin{figure}[ht]
    \centering
    \includegraphics[width=0.95\linewidth]{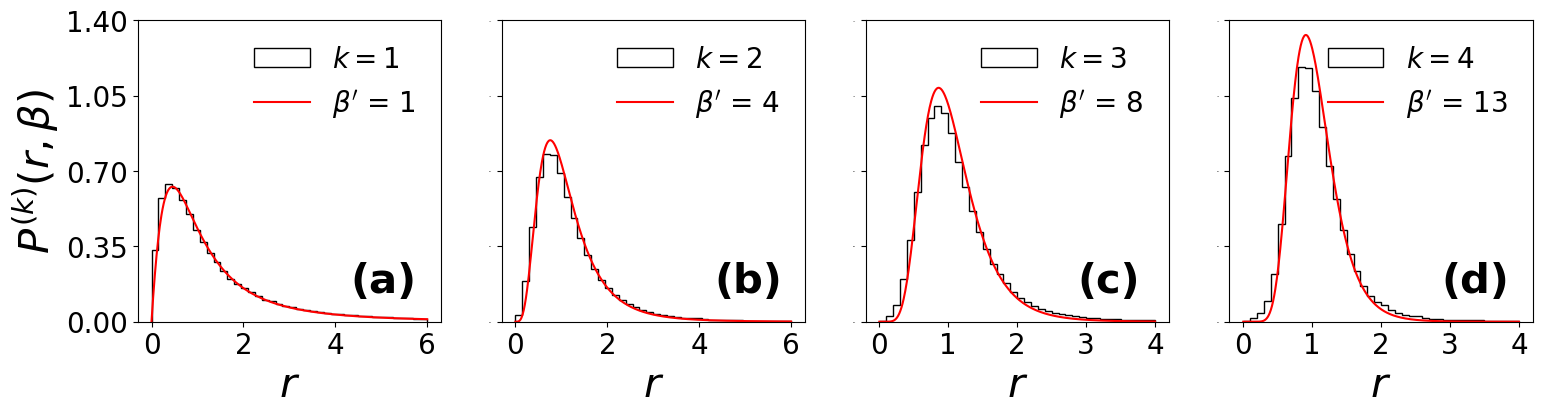}
    \caption{Plots for $k$-th order level spacing ratios [Eq.~\eqref{k-th-ratio_supp}] for the non-Hermitian Hamiltonian given in Eq.~\eqref{non-H-xxz_supp} with Hamiltonian given in Eq.~\eqref{eq:XXZ_supp} and the dissipative part given in Eq.~\eqref{eq:loss_supp}, with $\gamma$=2.0. Here (a)-(d) represents $k=1$ to $k=4$, respectively (black-solid). Similar to Fig.~\ref{fig:GinOE-NHRM_supp} we once again observe a good agreement with $\beta=1$ (red-solid). The results here are obtained for system size $M=16$ in the zero magnetization sector averaged over $50$ different realizations. In this case we find the KL divergences [Eq.~\eqref{KL-divergence_supp}] to be $0.0097$, $0.0262$, $0.0569$, and $0.0939$, for (a) -(d), respectively.}
    \label{fig:nhh1_supp}
\end{figure}

\begin{figure}[ht]
    \centering
    \includegraphics[width=0.95\linewidth]{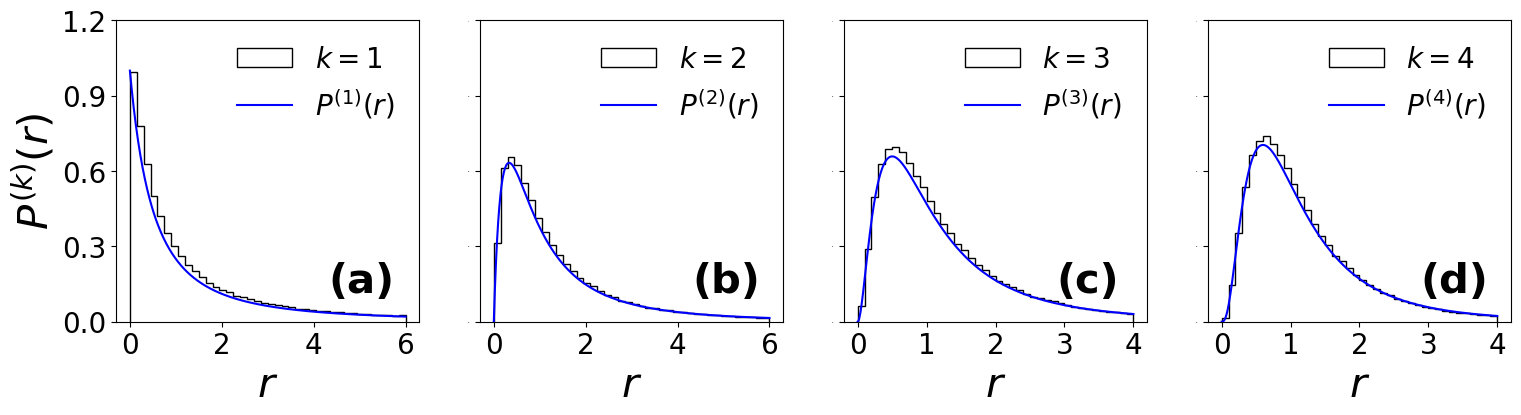}
    \caption{Plots for $k$-th order level spacing ratios [Eq.~\eqref{k-th-ratio_supp}] for the non-Hermitian Hamiltonian given in Eq.~\eqref{non-H-xxz_supp} with Hamiltonian given in Eq.~\eqref{eq:XXZ_supp} and the dissipative part given in Eq.~\eqref{eq:loss_supp}, with $\gamma=20$ which ensure that the system is in the localized regime. Here (a)-(d) represents $k=1$ to $k=4$, respectively (black-solid). Similar to Fig.~\ref{fig:Poisson_supp} we once again observe a good agreement with Poisson statistics (blue-solid). The results here are obtained for system size $M=16$ in the zero magnetization sector averaged over $50$ different realizations. The values of KL divergence are $0.010$, $0.0105$, $0.0031$ and $0.0039$ respectively.}
    \label{fig:nhh1p_supp}
\end{figure}

\section{XXZ model with dissipation and $k$-th level spacing ratio}
\label{sec:xxz}

In this section, we study singular value statistics of non-Hermitian version of XXZ Hamiltonian. In particular, we discuss an integrable interacting spin-$1/2$ Hermitian Hamiltonian consisting of $M$ qubit to which a disordered non-Hermitian contribution is added that describes random site-dependent losses \cite{chenu_svd}. The non-Hermitian Hamiltonian is of the form
\begin{equation}
    H  = H_\text{XXZ} - i \,  \Gamma/2\,,
    \label{non-H-xxz_supp}
\end{equation}
with the Hermitian part corresponding to the XXZ model given by
    \begin{equation}
         H_\text{XXZ}  = J \sum_{i=1}^M \left(  S_i^x  S_{i+1}^x +  S_i^y S_{i+1}^y + \Delta S_i^z  S_{i+1}^z  \right).
         \label{eq:XXZ_supp}
    \end{equation}
Here $S_i^{x,y,z}$ correspond to the spin operators in the $x$-, $y$- and $z$-directions, respectively and are given in terms of the Pauli matrices as $S^q = \frac{1}{2}\sigma^q, q=x,y,z$. Here $J$ is the nearest-neighbour hopping term is set to $1$ and we consider here the isotropic case by setting $\Delta=1$. The non-Hermiticity in Eq.~\eqref{non-H-xxz_supp} is explicitly added as a loss term $\Gamma$, which is given as 
    \begin{equation}
        \label{eq:loss_supp}
         \Gamma  = \sum_{i=1}^M \gamma_i \left(  S^z_i +1/2\right),
    \end{equation}
where $\gamma_i$ are rates independently sampled from a uniform distribution over the interval $[0, \gamma]$.  We assume periodic boundary condition and work in the zero magnetization sector. Note that the non-Hermitian Hamiltonian in Eq.~\eqref{non-H-xxz_supp} is very different than the non-Hermitian Hamiltonian discussed in the main text. In Eq.~\eqref{non-H-xxz_supp} the disorder is in the non-Hermitian term as opposed to the one considered in the main text. The system in Eq.~\eqref{non-H-xxz_supp} shows a transition in the ratios of the singular values from $\beta=1$ statistics to Poisson as one tunes from low $\gamma$ to high $\gamma$. This is shown in Fig.~\ref{fig:nhh1_supp} and Fig.~\ref{fig:nhh1p_supp}, respectively.

\end{document}